\documentclass[a4paper,12pt]{article}
\usepackage[pctex32]{graphics}
\usepackage{amssymb,amsmath}
\usepackage{amsmath,amssymb}
\usepackage{latexsym}
\usepackage{epsfig}
\usepackage[english]{babel}

\newcommand{\be}{\begin{equation}}
\newcommand{\ee}{\end{equation}}
\newcommand{\ba}{\begin{eqnarray}}
\newcommand{\ea}{\end{eqnarray}}

\begin{document}

\begin{titlepage}

\vspace{5mm}
\begin{center}

{\Large \bf Quasinormal frequencies using the hidden conformal
symmetry of the Schwarzschild black hole}

\vskip .6cm

\centerline{\large
 Yong-Wan Kim $^{1,a}$,  Yun Soo Myung$^{2,b}$,
and Young-Jai Park$^{1,3,c}$}

\vskip .6cm

{$^{1}$ Center for Quantum Spacetime,
Sogang University, Seoul 121-742, Korea}\\

{$^{2}$Institute of Basic Science and School of Computer Aided
Science, \\Inje University, Gimhae 621-749, Korea \\}

{$^{3}$Department of Physics and Department of Global Service Management,\\
Sogang University, Seoul 121-742, Korea}

\end{center}

\begin{center}
\underline{Abstract}
\end{center}

We show that the hidden conformal symmetry of the Schwarzschild
black hole is realized  from the AdS$_2$ sector of the AdS$_2\times
S^2$, but not from the Rindler spacetime which is the genuine
near-horizon geometry of the Schwarzschild black hole. This implies
that purely imaginary quasinormal frequencies obtained using the
hidden conformal symmetry is not suitable for describing the largely
damped modes around the Schwarzschild black hole.

\vskip .6cm


\vskip 0.8cm

\vspace{15pt} \baselineskip=18pt

\noindent $^a$ywkim65@gmail.com\\
\noindent $^b$ysmyung@inje.ac.kr \\
\noindent $^c$yjpark@sogang.ac.kr

\thispagestyle{empty}
\end{titlepage}

\newpage
\section{Introduction}
Recently, it was shown that an SL(2,R) hidden conformal
symmetry~\cite{Castro:2010fd} could be realized in a scalar wave
equation around the Schwarzschild black hole when taking the
near-region and low-energy limits~\cite{Bertini:2011ga}. We refer to
the geometry modified in this way as the {\it subtracted geometry}
which shows that it has the same near-horizon properties as the
original Schwarzschild black hole, but different asymptotes at
infinity~\cite{Cvetic:2011dn}.  Actually, it does not have an
asymptotically flat spacetime  implied by the Schwarzschild black
hole but it has an asymptotically anti-de Sitter (AdS) spacetime.

Importantly, it was claimed that a hidden conformal symmetry has
been used to derive purely imaginary quasinormal frequencies (QNFs)
by using the operator method.   However, we wish to clarify  that
the Rindler space is the truly near-horizon geometry of the
Schwarzschild black hole, while the near-region and low-energy
limits of  the wave equation around the Schwarzschild spacetime
corresponds to the wave equation around the AdS segment of the
AdS$_2\times S^2$,  which is inspired by the near-extremal
Reissner-Nordstr\"om (RN) black hole~\cite{Li:2011tga}.    Here the
near-horizon geometry (region) means $r\approx  r_+$, while both the
near-region ($\omega r \ll1$) and the low-energy limit ($\omega r_+
\ll 1$)~\cite{Maldacena:1997ih} are  necessary to develop a hidden
conformal symmetry in a scalar wave equation. The AdS segment
shrinks to the Rindler space for the non-extremal Schwarzschild
black hole, whereas it grows to the AdS space for the near-extremal
RN black hole.

On the other hand, it seems difficult to derive QNFs of a scalar
propagating on the Schwarzschild black hole by using a hidden
conformal symmetry solely.   As is well known, quasinormal modes
will be determined  by solving a scalar wave equation around the
Schwarzschild black hole as well as imposing the boundary
conditions: ingoing waves at the horizon and outgoing waves at
infinity.  We stress that   any restriction on the frequency
$\omega$ is not allowed for deriving quasinormal modes. Purely
imaginary QNFs were found by employing the operator method developed
in the {\it subtracted geometry}~\cite{Cvetic:2011dn}. In this case,
the ingoing waves at the horizon were guaranteed, but quasinormal
modes do not satisfy the outgoing wave-boundary condition  because
these modes were developed using the near-region and low-energy
limits~\cite{Bertini:2011ga} where the frequency $\omega$ should
satisfy inequalities of $\omega \ll 1/r$ and $\omega \ll 1/r_+$.
Furthermore, it was suggested that purely imaginary QNFs
($\omega_n$) in Eq.~(\ref{impq}) yield the correct leading behavior
of $\omega$ in Eq.~(\ref{nqn}) for large damping with large overtone
number $n$. However, we observe from Eq.~(\ref{impq}) that the
important differences are the absence of real part, the non-large
overtone number $n$, and the appearance of angular momentum number
$l$ in the imaginary part.  In view of these, purely imaginary QNFs
may not be acceptable as QNFs describing a scalar perturbation
around the Schwarzschild black hole because the outgoing boundary
condition is not imposed at infinity.

A promising case was found in the RN black hole.  We have derived
the purely imaginary QNFs of the RN black hole by making use of a
hidden conformal symmetry developed in the near-region and
low-energy limits of the scalar equation~\cite{Kim:2012mh}.   We are
aware that the operator approach has a limitation because the
$\omega$-dependent potential is approximated by a single term of a
hidden conformal symmetry potential. This means that developing a
hidden conformal symmetry in the near-horizon region  means
neglecting the large $r>r_+$ behavior of the potential in the whole
RN black hole spacetime and thus, leading to the {\it subtracted
geometry}. Fortunately, it was known that purely imaginary QNFs
could be also obtained from a scalar perturbation around the
near-extremal RN black hole without any
modifications~\cite{Chen:2012zn}.  This implies that  the {\it
subtracted geometry} [=near-extremal RN black hole] is the solution
to the original Einstein-Maxwell theory which provides the RN black
hole.

It is worth noting that both the near-horizon and asymptotic
geometries (whole information) are necessary to derive QNFs, even
though the entropy counting for a black hole may require the
near-horizon geometry only~\cite{Cvetic:2011dn,Compere:2012jk}. In
fact, it shows different utilities of the hidden conformal symmetry
in deriving between QNFs  and entropy. Hence we propose that if one
obtains QNFs (purely imaginary QNFs) using the hidden conformal
symmetry based on the {\it subtracted geometry}, one has to find the
corresponding black hole (near-extremal RN black hole) where the
same QNFs will be found by solving  the scalar equation as well as
imposing two boundary conditions without taking any limits.

One counter example is the  Schwarzschild black hole. In this case,
the {\it subtracted geometry} [=near-extremal RN black hole] is not
the solution to the original Einstein theory which provides the
Schwarzschild  black hole. We would like to stress  that the Rindler
spacetime is the near-horizon solution to the Einstein theory which
gives us the Schwarzschild black hole.

In this work, we will show that purely imaginary QNFs obtained using
a hidden conformal symmetry are not suitable for describing largely
damped modes around the Schwarzschild black hole.

\section{Hidden conformal symmetry in the subtracted geometry}

Let us begin with the Schwarzschild metric given by
 \be\label{Smetric}
 ds^2_{\rm Sch}=-f(r)dt^2+\frac{dr^2}{f(r)}+r^2d\Omega^2_2,
 \ee
where $f(r)=1-r_+/r$ with $r_+=2M$. Here, $M$ is the ADM mass and
the surface gravity is
 \be
 \kappa=\frac{1}{4M}=2\pi T_H
 \ee
with the Hawking temperature $T_H$.

The Klein-Gordon equation for a massless scalar is given by
 \be\label{SKG}
 \bar{\square}_{\rm Sch}\Phi = 0.
 \ee
Expanding in eigenmodes of
 \be\label{ansatz}
 \Phi(t,r,\theta,\phi) = e^{-i\omega t}\frac{R(r)}{r}Y^l_m(\theta,\phi)\ ,
 \ee
and using the tortoise coordinate defined by
 \be\label{tor}
 r_*=\int \frac{dr}{f(r)} = r+r_+\ln\left(\frac{r}{r_+}-1\right),
 \ee
the radial part of Eq.~(\ref{SKG}) becomes the Schr\"odinger-type
equation
 \be
  \frac{d^2}{dr_*^2} R(r) +\Big[\omega^2-V_{\rm Sch}(r)\Big]R(r)=0,
 \ee
where the potential is given by
 \be\label{pot}
 V_{\rm Sch}(r)=f(r)\Big[\frac{l(l+1)}{r^2}+\frac{2M}{r^3}\Big].
 \ee

On the other hand, the approximated Klein-Gordon equation could be
expressed in terms of  the eigenvalue
equation~\cite{Bertini:2011ga}
 \ba\label{hcs-sch}
 {\cal H}^2\Phi&=& \left(r^2f(r)\partial^2_r+2(r-M)\partial_r
                 -\frac{16M^4}{r^2f(r)}\partial^2_t\right)\Phi  \nonumber\\
 &=&l(l+1)\Phi
 \ea
which is designed  for describing a scalar propagating on the
Schwarzschild spacetime in the near-region and  low-energy limits.
Note that the SL(2,R) Casimir operator ${\cal H}^2$ is given by
 \be
 {\cal H}^2=-H^2_0+\frac{1}{2}\Big(H_1H_{-1}+H_{-1}H_1\Big),
 \ee
where three operators
 \ba
 H_{\pm 1} &=& \pm i e^{\pm t/4M}\Big(\Delta^{1/2}\partial_r
              \mp 4M(r-M)\Delta^{-1/2}\partial_t\Big),\nonumber\\
 H_0 &=& -4 i M \partial_t~~~{\rm with}~ \Delta=r^2f(r)
 \ea
obey the SL(2,R) commutation relations: \begin{equation}
\label{sl-alg} \Big[H_0,H_{\pm 1}\Big]=\mp i H_{\pm
1},~~\Big[H_1,H_{-1}\Big]=2iH_{0}.
\end{equation}

\begin{figure}[t!]
   \centering
   \includegraphics{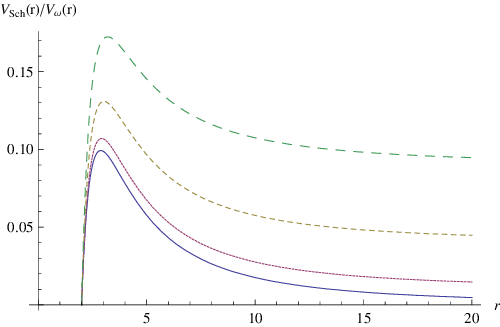}
\caption{Different potentials for $M=l=1$ in the Schwarzschild
coordinate $r$. The bottom curve depicts the Schwarzschild potential
$V_{\rm Sch}(r)$, while three others are the $\omega$-dependent
potential $V_\omega(r)$  for $\omega$=0.3, 0.2, and 0.1  from  top
to bottom. For large $r$, one observes that $V_{\rm Sch}(r) \sim 0$,
while $V_\omega(r)\sim \omega^2$. } \label{fig.1}
\end{figure}

In order to investigate what happens when  the near-region and
low-energy limits are taken into account, we rewrite the
approximated  Klein-Gordon equation (\ref{hcs-sch}) by using the
tortoise coordinate $r_*$ as
 \be
 \frac{d^2}{dr_*^2} R(r) +\Big[\omega^2-V_\omega(r)\Big]R(r)=0,
 \ee
where the  $\omega$-dependent potential
 \be
 V_\omega(r)=\omega^2\left(1-\frac{r^4_+}{r^4}\right)
          + V_{\rm Sch}(r)
 \ee
has an additional term to the original potential $V_{\rm Sch}(r)$.
Since the potential  $V_{\rm Sch}(r)$ contains all information for
the Schwarzschild black hole spacetime where a scalar propagates on,
the appearance of $\omega^2$-dependent term is unusual  and thus, it
reflects the near-region and low-energy limits.  We observe that the
$\omega^2$-dependent term arises from when replacing five terms in
$\omega^2r^4/\Delta$ with a single term
$\omega^2r^4_+/\Delta$~\cite{Bertini:2011ga}, which is regarded as a
key step to develop the hidden conformal symmetry in the
Schwarzschild spacetime.  This replacement  is done  when taking the
near-region limit ($\omega r \ll 1$) and low-energy limit ($\omega
r_+\ll 1$). If one does not take these  limits, the $r_+$ in the
$\omega^2$-dependent term goes back to $r$ and thus, the
$\omega^2$-dependent term disappears.  This implies that
$V_\omega(r)$ leads to $V_{\rm Sch}(r)$  for $r \to r_+$. In
Fig.~\ref{fig.1}, we depict the four potentials which consist of
$V_{\rm Sch}(r)$ and three $V_\omega(r)$ for $\omega=0.1,~0.2,~0.3$.
The figure shows that in the near-horizon limit ($r\to r_+=2$), all
the potentials have the nearly same form, irrespective of $\omega^2$
but for $r>r_+$, they have different forms depending on $\omega^2$.
Hence, we observe that for $r\approx r_+$, $V_\omega(r) \approx
V_{\rm Sch}(r) $ for any $\omega$. When $\omega$ approaches zero,
$V_\omega(r)$ recovers the original potential $V_{\rm Sch}(r)$.
However, it is worth noting that the near-horizon region of $r
\approx r_+$ is not enough to derive the quasinormal modes because
two boundary conditions are required. Definitely, one observes a
difference between $V_{\rm Sch}(r) \sim 0$ and $V_\omega(r)\sim
\omega^2$ for large $r$.

In order to describe the near-horizon region well, it is convenient
to introduce a new coordinate defined by
 \be\label{rho}
 \rho \equiv -\frac{1}{2\kappa}\ln\Big[1-\frac{r_+}{r}\Big].
 \ee
In terms of $\rho$, the event horizon $r_+$ is mapped into
$\rho\rightarrow\infty$, while the spatial infinity
$r\rightarrow\infty$ into $\rho\rightarrow 0$: $r \in[r_+,\infty]$
is inversely mapped into $\rho \in[\infty,0]$. Using the coordinate
(\ref{rho}), the Schwarzschild metric becomes
 \be\label{rho-metric}
 ds^2_{\rm \rho} =
  -\tilde{f}(\rho)dt^2
  +\tilde{f}^{-1}(\rho)\left(\frac{r_+}{2\sinh(\kappa \rho)}\right)^2
  \left[\left(\frac{\kappa}{\sinh(\kappa \rho)}\right)^2  d\rho^2
  +d\Omega^2\right]
 \ee
with
 \be
 \tilde{f}(\rho)=e^{-2\kappa\rho}.
 \ee
Then, making use of the ansatz
 \be
 \Phi(t,\rho,\theta,\phi) = e^{-i\omega t}R(\rho)Y^l_m(\theta,\phi)\ ,
 \ee
the massless scalar propagating in the spacetime (\ref{rho-metric})
satisfies
 \be
 \label{radial}
 \Big(\frac{\sinh(\kappa\rho)}{\kappa}\Big)^2\frac{d^2}{d\rho^2}R(\rho)
 +\left[\frac{\omega^2}{\tilde{f}^2(\rho)}\Big(\frac{r_+}{2\sinh(\kappa\rho)}\Big)^2
 - l(l+1)\right]R(\rho)=0.
 \ee
This can be rewritten as the Schr\"odinger-type equation
 \ba
 \label{sch2}
 \frac{d^2}{d\rho^2} R(\rho) +\Big[\omega^2-\tilde{V}_{\rm \omega}(\rho)\Big]R(\rho)=0.
 \ea
Here the $\omega$-dependent  potential is given by
 \be \label{pot2}
 \tilde{V}_{\rm \omega}(\rho)=
 \omega^2\Big[1-\frac{1}{16\tilde{f}^{2}(\rho)\sinh^4(\kappa\rho)} \Big]
 +\frac{l(l+1)\kappa^2}{\sinh^2(\kappa\rho)}.
 \ee
We note a useful relation between $r$ and $\rho$
 \be
 \left(\frac{r_+}{r}\right)^4
 =16\tilde{f}^{2}(\rho)\sinh^4(\kappa\rho).
 \ee
In the limits of the near-region ($\rho\rightarrow\infty$) and
low-energy ($\omega \to 0$),  the square bracket in (\ref{pot2}) is
approximated to be zero as
 \be
 \omega^2\Big[1-\frac{1}{16\tilde{f}^{2}(\rho)\sinh^4(\kappa\rho)}\Big]
 \sim \frac{\omega^2}{\sinh^2(\kappa\rho)} \sim 0.
 \ee
Thus, in these limits, the $\omega$-dependent potential
$\tilde{V}_\omega$ (\ref{pot2}) reduces to a single term
 \be\label{pot3}
  V_{\rm HCS}(\rho) =
  \frac{l(l+1)\kappa^2}{\sinh^2(\kappa\rho)}.
 \ee
We note that this potential is not  the original potential
$\tilde{V}_\omega$. This is just an approximated  potential to
explore the hidden conformal symmetry in the near-horizon region of
the Schwarzschild spacetime.  Hence, we have the {\it subtracted
geometry}  when using $V_{\rm HCS}$ instead of $\tilde{V}_\omega$.
Its  near-horizon and asymptotic forms take $V_{\rm HCS}(\rho\to
\infty)\sim e^{-2\kappa \rho}$ and $V_{\rm HCS}(\rho\to 0)\sim
\frac{1}{\rho^2}$, which show that $V_{\rm HCS}(\rho)$ is similar to
the potential of a scalar field around the AdS-black hole. This
suggests that  its asymptote is changed from a flat spacetime
implied by the Schwarzschild black hole to an AdS spacetime.

In order to exhibit the hidden conformal structure, we construct
three vectors defined as
 \ba
 H_{1/-1}&=&\frac{i}{\kappa}e^{\pm\kappa t}
        \Big[\cosh(\kappa\rho)\partial_t\pm\sinh(\kappa\rho)\partial_\rho\Big],\nonumber\\
 H_0 &=& - \frac{i}{\kappa}\partial_t,
 \ea
which satisfy the SL(2,R) algebra  (\ref{sl-alg}). Then, the
SL(2,R) Casimir operator is given by
 \be\label{hcs-sch1}
 {\cal H}^2=-\Big[\frac{\sinh(\kappa\rho)}{\kappa}\Big]^2\partial^2_t
              +\Big[\frac{\sinh(\kappa\rho)}{\kappa}\Big]^2\partial^2_\rho.
 \ee
As a result, the Schr\"odinger equation (\ref{sch2}) with $V_{\rm
HCS}(\rho)$ in Eq.~(\ref{pot3}) instead of $\tilde{V}_\omega(\rho)$
is equivalent to the eigenvalue equation
 \be\label{eigenv}
 {\cal H}^2\Phi=l(l+1)\Phi.
 \ee

Now, one may use the hidden conformal symmetry to derive QNFs  of
the Schwarzschild black hole represented by the {\it subtracted
geometry}. We start with the primary state given by
 \be
 H_0\Phi^{(0)}=ih\Phi^{(0)}
 \ee
which satisfies   the highest weight condition \be \label{hwc}
 H_1\Phi^{(0)}=0. \ee
Then, using the ansatz
 \be
 \Phi^{(0)}=e^{-i\omega_0 t}R^{(0)}(\rho)Y^l_m(\theta,\phi),
 \ee
one has a conformal weight
 \be\label{h1}
 h=i\frac{\omega_0 }{\kappa}=i \frac{\omega_0}{2\pi T_H}.
 \ee
On the other hand,  the SL(2,R) Casimir operator satisfies
 \be \label{caop}
 {\cal H}^2\Phi^{(0)}=h(h+1)\Phi^{(0)}.
 \ee
Comparing Eq. (\ref{caop}) with Eq. (\ref{eigenv}), one has
 \be
 h=\frac{1}{2}[1\pm(2l+1)],
 \ee
and thus, one finds
 \be
 \omega_0=-i\frac{\kappa}{2}[1\pm(2l+1)].
 \ee
Since the QNFs are purely imaginary
$\omega_I>0~(\omega=\omega_R-i\omega_I)$ with $\omega_R=0$, we
choose the upper sign as
 \be
 \omega_0=-i\kappa(l+1).
 \ee
All the descendants can be constructed  by acting the operator
$H_{-1}$ on $\Phi^{(0)}$
 \be
 \Phi^{(n)}=(-iH_{-1})^n\Phi^{(0)}
 \ee
so that we have
 \be
 \Phi^{(n)}=e^{-i\omega_n t}R^{(n)}(\rho)Y^l_m(\theta,\phi).
 \ee
Here the QNFs could be  read off as
 \be \label{impq}
 \omega_n=\omega_0-i\kappa n=-i\kappa\Big[n+l+1\Big],~~~n\in Z^+
 \ee
which are purely imaginary.

On the other hand, numerical computations of the QNFs for the
Schwarzschild black hole in the limit of large damping is given
by~\cite{Nollert:1993zz,Motl:2002hd}
 \be \label{nqn}
 \omega= (1.098612\cdots)
 T_{H}-i 2\pi  T_{H} \Big(n+\frac{1}{2}\Big),
 \ee
where the real part approaches a constant of
$\ln(3)$~\cite{Hod:1998vk}, while the imaginary part becomes equally
spaced with large $n$ ($\omega_I\gg1$). Comparing (\ref{impq}) with
the highly damped  frequencies (\ref{nqn}), one may see apparently
that the leading behavior of $i2\pi T_H n$ for large damping comes
out correctly. However, we observed from (\ref{nqn}) that the
important differences are the absence of real part, the non-large
$n$, and the appearance of $l$ in the imaginary part. These show
that the QNFs $\omega_n$ are not suitable for describing the scalar
perturbation absorbing into the black hole.

Moreover, the $n$-th radial eigenfunction $ R^{(n)}(\rho)$ is
constructed as
 \ba
 R^{(n)}(\rho)&=&\left(\kappa\right)^{-n}
    \left(-i\omega_{n-1}\cosh\left(\kappa\rho\right)-\sinh\left(\kappa\rho\right)\frac{d}{d\rho}\right)\nonumber\\
    &&~\times\left(-i\omega_{n-2}\cosh\left(\kappa\rho\right)-\sinh\left(\kappa\rho\right)\frac{d}{d\rho}\right)\nonumber\\
    &&~\cdot\cdot\cdot \times \left(-i\omega_0\cosh\left(\kappa\rho\right)-\sinh\left(\kappa
    \rho\right)\frac{d}{d\rho}\right)R^{(0)}(\rho),
 \ea
 which may be  regarded as the $n$-th radial  quasinormal modes.
We note that $\Phi^{(n)}$   forms a principally  discrete highest
weight representation of the SL(2,R)
 \be
 H_0\Phi^{(n)}=i(h+n)\Phi^{(n)}.
 \ee
On the other hand, the highest weight condition (\ref{hwc}) provides
the radial solution
 \be\label{R0sol}
 R^{(0)}(\rho)=C\Big[\sinh\left(\kappa\rho\right)\Big]^{i\frac{\omega_0}{\kappa}}.
 \ee
Near the horizon of $\rho\to \infty~(r_*\to -\infty)$, the solution
(\ref{R0sol}) behaves as
 \be
 R^{(0)}(\rho) \sim e^{i\omega_0 \rho},
 \ee
showing that this is the outgoing mode ($\rightarrow$) into the
horizon which is equivalent to the ingoing mode ($\leftarrow$) at
the horizon when using the tortoise coordinate $r_*$ in
Eq.~(\ref{tor}). We observe that $R^{(0)}(0)\sim 0$, which may show
that it is  not  the outgoing wave at infinity but it satisfies the
Dirichlet boundary condition at the infinity of AdS spacetime.

Furthermore, the first radial eigenfunction $R^{(1)}(\rho)$ can be
explicitly constructed as
 \be
 R^{(1)}(\rho) =
 \begin{cases}
  -2iC\omega_0\cosh(\kappa\rho)\Big[\sinh\left(\kappa\rho\right)\Big]^{i\frac{\omega_0}{\kappa}}
            & \sim~~0~~\,{\rm as~}\rho\rightarrow 0, \\
 -2iC\omega_0\coth(\kappa\rho)\Big[\sinh\left(\kappa\rho\right)\Big]^{i\frac{\omega_1}{\kappa}}
            & \sim~~ e^{i\omega_1 \rho}~~{\rm as~}\rho\rightarrow \infty,
 \end{cases}
 \ee
which show that at infinity $R^{(1)}(\rho)$ satisfies the Dirichlet
boundary condition, while near the horizon $R^{(1)}(\rho)$ remains
to be the outgoing mode. One can easily show that the $n$-th radial
eigenfunction $ R^{(n)}(\rho)$ behaves as the same way as
$R^{(1)}(\rho)$ by induction.

In order to obtain the truly QNFs, we  have to impose the two
boundary conditions: outgoing mode  near the horizon ($\rho \to
\infty$) and ingoing mode  at infinity ($\rho \to 0$). However, we
point out that $R^{(n)}(\rho)$ do not satisfy the ingoing boundary
condition  at infinity   because they are the solutions which were
obtained by considering the {\it subtracted geometry}.  Curiously,
we observe that $R^{(n)}(\rho)\sim 0$ as $\rho\rightarrow 0$, which
implies that in this framework, one could not impose the ingoing
boundary condition at infinity. In this sense, we could not regard
(\ref{impq}) as the truly QNFs which describe the largely damped
modes around the Schwarzschild black hole  because we have
considered the {\it subtracted geometry}.

\section{QNFs of scalar  around  AdS-segment}

In this section, we show explicitly that the approximated potential
(\ref{pot3}) comes from the AdS$_2$-segment inspired by the fact
that AdS$_2\times S^2$ is the near-extremal RN black hole, but not
from the Rindler space  which is the genuine near-horizon geometry
of the Schwarzschild spacetime. This implies that we will no longer
use the {\it subtracted geometry} to derive the QNFs of scalar
around the Schwarzschild black hole.

First of all, we mention that the near-horizon geometry of the
Schwarzschild spacetime is a product of the Rindler space and a two
sphere $S^2$.  This can be easily realized  when using  the
coordinate transformations
 \be\label{nearS}
 \hat{t}=\frac{t}{2r_+},~~~r=r_+\left(1+\frac{\eta^2}{4}\right).
 \ee
The Schwarzschild spacetime (\ref{Smetric}) is turned out to be the
Rindler spacetime
 \be\label{Snear}
 ds^2_{\rm RS}=r^2_+\Big(-\eta^2d\hat{t}^2+d\eta^2+d\Omega^2_2\Big)
 \ee
in the near-horizon region. Note that it is impossible to develop
a hidden conformal symmetry in the Rindler spacetime as will be
shown in the next section.

For reference, we introduce the RN spacetime described by
\be\label{RN}
 ds^2_{\rm RN}=
  -f(r)dt^2+f^{-1}(r)dr^2+r^2d\theta^2+r^2 d\Omega^2_2,
 \ee
where the metric function with mass $M$ and charge $Q$ is given by
 \be
 f(r)=1-\frac{2M}{r}+\frac{Q^2}{r^2}.
 \ee
The inner ($r_-$) and the outer ($r_+$) horizons are obtained as
 \be
 r_\pm=M\pm\sqrt{M^2-Q^2}\equiv M\pm r_0,
 \ee
which satisfy $f(r_\pm)=0$. We note that $r_0$ is a non-extremal
parameter, but a very small $r_0 \ll M(\sim Q)$ corresponds to the
near-extremal RN black hole.  Also,  we have an extremal RN black
hole for $r_0=0$.

On the other hand, the
AdS$_2$-segment~\cite{Li:2011tga,Carroll:2009maa} is introduced to
be
 \be\label{ads2}
 ds^2_{\rm AdS}=Q^2\Big(-\sinh^2\!\!\eta\, d\hat{t}^2
                +\frac{\cosh^2\!\!\eta}{\sinh^2\!\!\eta+r^2_0}d\eta^2+d\Omega^2_2\Big),
 \ee
 which describes the near-extremal RN black hole.
We note here that for $r_0=1$~\cite{Li:2011tga}, (\ref{ads2})
reduces to the near-horizon geometry (AdS$_2\times S^2$,
Bertotti-Robinson spacetime) of the extremal RN black
hole~\cite{Matyjasek:2004gh,Myung:2007an} \be\label{ads2li}
 ds^2_{\rm ERN}=Q^2\Big(-\sinh^2\!\!\eta\, d\hat{t}^2
                +d\eta^2+d\Omega^2_2\Big),
 \ee
 which is a solution to the Einstein-Maxwell theory.
 In this extremal case, its Hawking temperature is zero and its
QNFs are not defined properly~\cite{Chen:2012zn}. The spacetime is
described by  Ricci scalars
 \be
 R_{AdS_2}=-\frac{2}{Q^2},~~R_{S^2}=\frac{2}{Q^2},~~R=R_{AdS_2}+R_{S^2}=0.
 \ee
It is clear that  (\ref{ads2li}) could not describe the
near-extremal RN black hole because it contains only $Q$ when
comparing to (\ref{ads2}) with two parameters $Q$ and $r_0$.

 We are interested in obtaining QNFs of scalar
around the black hole with non-zero Hawking temperature. Hence, we
require $r_0 \not=1$. In fact, in the limit of $\eta\rightarrow 0$
together with $r_0= 1$ and $Q=r_+$, (\ref{ads2}) reduces to the
Rindler metric (\ref{Snear}). In order to make a further connection
to the near-extremal RN black hole, we introduce the coordinate
transformations
 \be
 \sinh^2\!\!\eta=\tilde\rho^2-r^2_0,~~~\hat{t}=\frac{t}{Q^2}.
 \ee
Then, one can obtain the near-extremal RN black hole with
$r_0=(r_+-r_-)/2\ll1$
 \be\label{nearRN}
 ds^2_{\rm
 NERN}=-\frac{\tilde\rho^2-r^2_0}{Q^2}dt^2+\frac{Q^2}{\tilde\rho^2-r^2_0}d\rho^2+Q^2d\Omega^2_2
 \ee
from the metric (\ref{ads2}).  Here $\tilde\rho\in[r_0,\infty]$. For
this near-extremal RN black hole, the surface gravity is computed to
be
 \be \label{dikappa}
 \tilde\kappa=\frac{r_0}{Q^2}.
 \ee
Introducing the tortoise coordinate defined by
 \be
 \rho_*=\frac{1}{2\tilde\kappa}\ln\left(\frac{\tilde{\rho}+r_0}{\tilde{\rho}-r_0}\right),~~\rho_*\in
 [\infty,0],
 \ee
the Klein-Gordon equation for the radial coordinate becomes the
Schr\"odinger-type equation
 \be
 \frac{d^2}{d\rho^2_*}R(\rho_*) + \Big[\omega^2-V_{\rm NERN}(\rho_*)\Big]R(\rho_*)=0,
 \ee
where the near-extremal RN potential is given by
 \be \label{neblp}
 V_{\rm NERN}(\rho_*)=\frac{l(l+1)\tilde{\kappa}^2}{\sinh^2(\tilde{\kappa} \rho_*)}.
 \ee
This becomes the  same form of $V_{\rm HCS}(\rho)$ in Eq.
(\ref{pot3}) when replacing $\tilde{\kappa}$ and $\rho_*$ by
$\kappa$ and $\rho$. An important fact to point out is that the
near-extremal RN potential is valid for whole spacetimes outside the
horizon because we consider the near-extremal RN black hole itself.

Next, in order to exhibit a hidden conformal structure of the
spacetime (\ref{nearRN}), we construct three vectors as
 \ba\label{hcs-ads}
 \hat{H}_{1/-1}&=& \frac{i}{\tilde\kappa} e^{\pm\tilde{\kappa}t}
        \Big(\frac{\tilde\rho}{(\tilde\rho^2-r^2_0)^{1/2}}\partial_t
        \mp\tilde\kappa(\tilde\rho^2-r^2_0)^{1/2}\partial_{\tilde\rho}\Big),\nonumber\\
 \hat{H}_0 &=& - \frac{i}{\tilde\kappa}\partial_t
 \ea
satisfying the SL(2,R) algebra (\ref{sl-alg}). Then, the SL(2,R)
Casimir operator is given by
 \be\label{casimir1}
 \hat{\cal H}^2=
 \Big[(\tilde\rho^2-r^2_0)^{1/2}\partial^2_{\tilde\rho}
 +2\tilde\rho\partial_{\tilde\rho}-\frac{Q^4}{(\tilde\rho^2-r^2_0)}\partial^2_t\Big].
 \ee
As a result, the radial equation can be rewritten as the eigenvalue
equation
 \be\label{eigenv-ads}
 \hat{\cal H}^2\Phi=l(l+1)\Phi,
 \ee
which is valid for whole space of  $\tilde\rho \in [r_0,\infty]$.
Here, using the hidden conformal symmetry (\ref{hcs-ads}) as before,
we can easily find the QNFs as
 \be \label{qnfa}
 \omega_n=\omega_0-i \tilde\kappa n=-i\tilde\kappa\Big[n+l+1\Big]
 \ee
with different surface gravity $\tilde{\kappa}$ in (\ref{dikappa}).

At this stage, we wish to point out that  purely imaginary QNFs
(\ref{qnfa}) exist as quasinormal frequencies  around the
near-extremal RN black hole  whose geometry is AdS$_2\times S^2$ in
(\ref{nearRN}), but not asymptotically flat spacetime in (\ref{RN}).
While (\ref{qnfa}) is the same with the result (\ref{impq}), these
based on the hidden conformal symmetry are different from the known
results due to different boundary conditions: numerical computations
for the QNFs for the RN black hole in the extremal limit have shown
that there is the non-vanishing real part~\cite{Onozawa:1995vu}. On
the other hand, in the limit of large damping ($\omega_I\gg1$), the
extremal RN black hole has vanishing real part~\cite{Cho:2005yc}.
Any derivation from the extremal RN black hole provides again
non-zero real part for $\omega_I \gg 1$.
 In
particular, both were obtained under the boundary conditions:
ingoing mode at horizon and outgoing mode at infinity (see the
review article~\cite{Berti:2009kk}, in detail).  In this respect,
our pure imaginary QNFs (\ref{qnfa}) to near-extremal RN black hole
are not appropriate as a solution that satisfies asymptotically flat
boundary condition, but appropriate as a solution that satisfies the
Dirichlet boundary condition at the spatial infinity.

The $n$-th radial eigenfunction $ R^{(n)}(\tilde{\rho})$ is now
obtained as
 \ba
 R^{(n)}(\tilde\rho)&=&
   \tilde\kappa^{-n} \left(-\frac{i\omega_{n-1}\tilde\rho}{(\tilde\rho^2-r^2_0)^{1/2}}
   +\tilde\kappa(\tilde\rho^2-r^2_0)^{1/2}\frac{d}{d\tilde\rho}\right)\times
   \nonumber\\
   && \left(-\frac{i\omega_{n-2}\tilde\rho}{(\tilde\rho^2-r^2_0)^{1/2}}
   +\tilde\kappa(\tilde\rho^2-r^2_0)^{1/2}\frac{d}{d\tilde\rho}\right)\times \cdot\cdot\cdot
   \nonumber\\
   &&
    \left(-\frac{i\omega_0\tilde\rho}{(\tilde\rho^2-r^2_0)^{1/2}}
    +\tilde\kappa(\tilde\rho^2-r^2_0)^{1/2}\frac{d}{d\tilde\rho}\right)R^{(0)}(\tilde\rho).
 \ea
On the other hand, by solving the highest weight condition
 \be
 \hat{H}_1\Phi^{(0)}=0,
 \ee
we have
 \be
 R^{(0)}(\tilde\rho)=D\Big(\tilde\rho^2-r^2_0\Big)^{-\frac{i\omega_0}{2\tilde\kappa}}.
 \ee
In terms of $\rho_*$, it can be expressed as \be
 R^{(0)}(\rho_*)=D \Big[ \sinh(\tilde{\kappa} \rho_*)\Big]^{i \frac{\omega_0}{\tilde{\kappa}}}.
 \ee
Near the horizon of $\rho_*\to \infty~(\tilde\rho\rightarrow r_0)$,
it behaves as
 \be
  R^{(0)}(\rho_*)\sim e^{i\omega_0\rho_*},
 \ee
which is  the outgoing mode ($\rightarrow$) into the horizon. At
infinity ($\rho_* \to 0,~\tilde\rho\rightarrow\infty$), we observe
that
 \be
  R^{(0)}(\rho_*)\sim 0,
 \ee
which satisfies  the Dirichlet  boundary condition at the spatial
infinity of the AdS$_2$ spacetime.

Furthermore, the first radial eigenfunction $R^{(1)}(\tilde\rho)$
can be explicitly constructed as
 \be
 R^{(1)}(\tilde\rho) = -\frac{2iD\omega_0\tilde\rho}{\tilde\kappa}
            \Big(\tilde\rho^2-r^2_0\Big)^{-\frac{i\omega_0+\tilde\kappa}{2\tilde\kappa}}.
  \ee
Expressing it in terms of $\rho_*$ and then,  checking the boundary
conditions at both sides leads to \be
 R^{(1)}(\rho_*) =
 \begin{cases}
  -2iD\omega_0\cosh(\tilde{\kappa}\rho_*)\Big[\sinh\left(\tilde{\kappa}\rho_*\right)\Big]^{i\frac{\omega_0}{\tilde{\kappa}}}
            & \sim~~0~~~\,{\rm as~}\rho_*\rightarrow 0, \\
 -2iD\omega_0\coth(\tilde{\kappa}\rho_*)\Big[\sinh\left(\tilde{\kappa}\rho_*\right)\Big]^{i\frac{\omega_1}{\tilde{\kappa}}}
            & \sim~~ e^{i\omega_1 \rho_*}~~{\rm as~}\rho_*\rightarrow
            \infty.
 \end{cases}
 \ee
It shows that at infinity $R^{(1)}(\rho_*)$ satisfies the Dirichlet
boundary condition, while near the horizon, $R^{(1)}(\rho_*)$
remains to be the outgoing mode. One can easily show that the $n$-th
radial eigenfunction $ R^{(n)}(\rho_*)$ satisfies  the same
AdS-boundary condition as $R^{(1)}(\rho_*)$ by induction.

\section{QNFs in the Rindler spacetime}

First of all, we investigate the boundary conditions for QNFs in the
Rindler spacetime. For that purpose, let us introduce the tortoise
coordinate
 \be
 u=-\frac{1}{\kappa}\ln\eta,
 \ee
which maps $\eta\in [0,\infty]$ inversely  into $u\in [\infty,0]$.
The Schr\"odinger-type equation can be written as
 \be
 \frac{d^2}{du^2} R+ \big[\omega^2 -V_{\rm RS}(u)\Big] R=0. \ee
Here the Rindler potential is given by
 \be
 V_{\rm RS}(u)= l(l+1)\kappa^2 e^{-2\kappa u},
 \ee
which is not the form of the potential $V_{\rm HCS}(\rho)$ in
Eq.~(\ref{pot3}) when replacing $u$ by $\rho$.  Near the horizon
($u\rightarrow \infty$), the Rindler potential shows the same
behavior as $V_{\rm AdS}(\rho)$, while it shows different behavior
at infinity ($u \to 0$). Thus, for quasinormal mode-boundary
condition, one requires an outgoing mode as $R(u)\sim e^{i\omega u}$
(ingoing mode $R(\eta)\sim e^{-i\omega \eta}$ in terms of $\eta$)
near the horizon. Since the potential height is finite as $l(l+1)
\kappa^2$ at infinity ($u\rightarrow 0)$, one may  require the
ingoing mode (outgoing mode expressed in terms of $\eta$).

However, it turned out that there is no QNFs in the Rindler
spacetime satisfying such QNFs boundary
conditions~\cite{Natario:2004jd} because the modes satisfying near
horizon boundary condition fail to fulfill the boundary condition at
infinity at the same time.

Moreover, for the Rindler spacetime which is the genuine
near-horizon limit of the Schwarzschild black hole spacetime, the
SL(2,R) algebra is not closed.  Explicitly, three vectors obtained
from Eq.~(\ref{hcs-ads}) in the small $\eta$ limit as
 \ba
 H'_{1/-1}= -i e^{\pm\hat{t}}
        \Big(\frac{1}{\eta}\hat\partial_t\mp\partial_\eta\Big),~~
 H'_0 =- i\hat\partial_t
 \ea
give the following commutation relation
 \be
 [H'_0,H'_{\pm1}]=\mp iH'_{\pm1},~~[H'_1,H'_{-1}] =0,
 \ee
whose last term   does not satisfy the SL(2,R) algebra
(\ref{sl-alg}).

\section{Conclusion}
We have shown that the near-horizon limit of the Schwarzschild black
hole is the Rindler spacetime but not the near-region and low-energy
limits of the Schwarzschild black hole known as the {\it subtracted
geometry}. It was shown that the hidden conformal symmetry is
developed for the {\it subtracted geometry} only, but the {\it
subtracted geometry} is not sufficient  to derive the QNFs.

The purely imaginary QNFs (\ref{impq}) developed on the {\it
subtracted geometry} without imposing outgoing boundary condition at
infinity   have also been obtained as (\ref{qnfa}) from the
quasinormal modes which satisfy two boundary conditions of AdS$_2$
spacetime: ingoing mode at the horizon and Dirichlet boundary
condition at infinity. The latter boundary condition is designed for
the near-extremal RN black hole and thus, is sharply contrasted to
the outgoing mode-boundary condition imposed at asymptotically flat
spacetime of the Schwarzschild (RN) black hole.

 Hence, purely imaginary QNFs obtained from  the
{\it subtracted geometry}  could not describe the largely damped
modes of scalar perturbation around the Schwarzschild black hole.
When comparing (\ref{impq}) with (\ref{nqn}),
 we have observed that the important differences
are the absence of real part, the non-large overtone number $n$, and
the appearance of angular momentum number $l$ in the imaginary part.
This shows clearly  that the QNFs $\omega_n$ (\ref{impq})  are not
suitable for describing the scalar perturbation absorbing into the
Schwarzschild black hole with temperature $T_H=1/8\pi M$. The purely
imaginary QNFs $\omega_n$ is suitable for the scalar absorbing  into
the near-extremal RN black hole with temperature
$\tilde{T}_H=r_0/2\pi Q^2$.

Finally, we wish to comment on the disappearance of real part
($\omega_R=0$)  of QNFs in the hidden conformal symmetry
approximation of RN black hole (=near-extremal RN black hole).
Numerical computations of the QNFs for the extremal RN black hole
have shown that there is the non-vanishing real
part~\cite{Onozawa:1995vu}. However, in the limit of large damping
($\omega_I\gg1$), the extremal RN black hole has vanishing real
part~\cite{Cho:2005yc}. Any derivation from the extremal RN black
hole provides again non-zero real part for $\omega_I \gg 1$. We
stress here that both were obtained under the boundary conditions:
ingoing mode at horizon and outgoing mode at
infinity~~\cite{Berti:2009kk}. However,  our pure imaginary QNFs
(\ref{qnfa}) have been obtained based on boundary conditions:
ingoing mode at  horizon and the Dirichlet boundary condition at the
spatial infinity. Thus, the difference arises from the different
asymptotic geometry: our asymptotic spacetime of AdS$_2\times S^2$
is the near-horizon geometry of near-extremal RN black hole, while
the full geometry of near-extremal RN black hole is asymptotically
flat. This difference between near-horizon geometry  and the full
geometry makes a big difference in the spectrum of QNFs because this
gave arise to different boundary condition at infinity.

Hence, it is fair to say that  even for the near-extremal RN black
hole, its hidden conformal symmetry approximation (=its near-horizon
approximation) could not capture  the  QNFs of the full geometry of
the near-extremal RN black hole.

\section*{Acknowledgement}
This work was supported by the National Research Foundation of Korea
(NRF) grant funded by the Korea government (MEST) through the Center
for Quantum Spacetime (CQUeST) of Sogang University with grant
number 2005-0049409. Y. S. Myung was also supported by the National
Research Foundation of Korea (NRF) grant funded by the Korea
government (MEST) (No.2012-040499). Y.-J. Park was also supported by
World Class University program funded by the Ministry of Education,
Science and Technology through the National Research Foundation of
Korea(No. R31-20002).


\begin{thebibliography}{99}

\bibitem{Castro:2010fd}
  A.~Castro, A.~Maloney and A.~Strominger,
   Phys.\ Rev.\ D {\bf 82}, 024008 (2010)  [arXiv:1004.0996 [hep-th]].

\bibitem{Bertini:2011ga}
  S.~Bertini, S.~L.~Cacciatori and D.~Klemm,
  Phys.\ Rev.\ D {\bf 85}, 064018 (2012)  [arXiv:1106.0999 [hep-th]].


\bibitem{Cvetic:2011dn}
  M.~Cvetic and F.~Larsen,
  JHEP {\bf 1209}, 076 (2012)  [arXiv:1112.4846 [hep-th]].


\bibitem{Li:2011tga}
  H.~Li,
   arXiv:1108.0220 [hep-th].

\bibitem{Maldacena:1997ih}
  J.~M.~Maldacena and A.~Strominger,
  Phys.\ Rev.\ D {\bf 56}, 4975 (1997)  [hep-th/9702015].  

\bibitem{Kim:2012mh}
  Y.~-W.~Kim, Y.~S.~Myung and Y.~-J.~Park,
   arXiv:1205.3701 [hep-th].

\bibitem{Chen:2012zn}
  C.~-M.~Chen, S.~P.~Kim, I-C.~Lin, J.~-R.~Sun and M.~-F.~Wu,
   Phys.\ Rev.\ D {\bf 85}, 124041 (2012)  [arXiv:1202.3224 [hep-th]].

\bibitem{Compere:2012jk}
  G.~Compere,
   arXiv:1203.3561 [hep-th].

\bibitem{Nollert:1993zz}
  H.~-P.~Nollert,
  Phys.\ Rev.\ D {\bf 47}, 5253 (1993).  


\bibitem{Motl:2002hd}
  L.~Motl,
   Adv.\ Theor.\ Math.\ Phys.\  {\bf 6}, 1135 (2003)  [gr-qc/0212096].

\bibitem{Hod:1998vk}
  S.~Hod,
   Phys.\ Rev.\ Lett.\  {\bf 81}, 4293 (1998)  [gr-qc/9812002].

\bibitem{Carroll:2009maa}
  S.~M.~Carroll, M.~C.~Johnson and L.~Randall,
  JHEP {\bf 0911}, 109 (2009)  [arXiv:0901.0931 [hep-th]].

\bibitem{Matyjasek:2004gh}
  J.~Matyjasek,
  Phys.\ Rev.\ D {\bf 70}, 047504 (2004)  [gr-qc/0403109].
\bibitem{Myung:2007an}
  Y.~S.~Myung, Y.~-W.~Kim and Y.~-J.~Park,
  \ Phys.\ Rev.\ D {\bf 76}, 104045 (2007)  [arXiv:0707.1933 [hep-th]].

\bibitem{Onozawa:1995vu}
  H.~Onozawa, T.~Mishima, T.~Okamura and H.~Ishihara,
  Phys.\ Rev.\ D {\bf 53}, 7033 (1996)  [gr-qc/9603021].

\bibitem{Cho:2005yc}
  H.~T.~Cho,
  Phys.\ Rev.\ D {\bf 73}, 024019 (2006)  [gr-qc/0512052].

\bibitem{Berti:2009kk}
  E.~Berti, V.~Cardoso and A.~O.~Starinets,
  Class.\ Quant.\ Grav.\  {\bf 26}, 163001 (2009)  [arXiv:0905.2975 [gr-qc]].


\bibitem{Natario:2004jd}
  J.~Natario and R.~Schiappa,
  Adv.\ Theor.\ Math.\ Phys.\  {\bf 8}, 1001 (2004)  [hep-th/0411267].


 \end{thebibliography}
\end{document}